\documentclass[12pt]{article}
\newcommand{\thetab}{{\bar\theta}}

\newcommand{\beq}{\begin{eqnarray}}
\newcommand{\eeq}{\end{eqnarray}}
\newcommand{\nn}{\nonumber}
\def\){\right)}
\def\({\left( }
\def\]{\right] }
\def\[{\left[ }
\def\CL{{\cal L}}
\def\CM{{\cal M}}
\def\mup{{m_u}}
\def\mdown{{m_d}}
\def\ms{{m_s}}
\def\mh{{\hat{m}}}
\def\mt{{\bar{m}}}
\def\msea{{m_{\rm sea}}}

\def\diag{{\rm diag}}
\def\tr{{\rm\ Tr}}

\title{Testing $\mup=0$ on the Lattice}

\author{Andrew G. Cohen$^a$, David B. Kaplan$^b$, and
  Ann E. Nelson$^c$\ \thanks{\tt cohen@bu.edu,
    dbkaplan@phys.washington.edu, anelson@fermi.phys.washington.edu}\\ \\
  \small \sl $^a$Department of Physics, Boston University, Boston, MA
  02215, USA\\ \\
  \small \sl $^b$ Institute for Nuclear Theory, Box 1550,
           University of Washington, Seattle, WA 98195-1550, USA \\ \\
  \small \sl $^c$Department of Physics, Box 1560, University of Washington,
           Seattle, WA 98195-1560, USA\\ \\
  }

\begin{document} 
\setlength{\baselineskip}{24pt}
\begin{titlepage}
\maketitle
\begin{picture}(0,0)(0,0)
\put(295,370){BUHEP-99-22}
\put(295,360){DOE/ER/40561-69-INT99}
\put(295,350){UW/PT-99/23}
\end{picture}
\vspace{-36pt}


\begin{abstract}
A massless up quark is an intriguing possible solution to the strong CP
problem.
We discuss how  lattice computations can be used in conjunction with
chiral perturbation theory to address the consistency of $\mup=0$ 
 with the observed hadron spectrum and
interactions.  It is not necessary to simulate very light
quarks---three flavor  partially quenched computations with comparable sea and
valence quark masses on the order of the strange quark mass could suffice.
\end{abstract}
\thispagestyle{empty}
\setcounter{page}{0}
\end{titlepage}

\section{Introduction}
QCD allows violation of the symmetry CP through the parameter
$\thetab\equiv \theta_{QCD}-{\rm arg\ det }{\CM}$, where $\CM$ is the quark
mass matrix. All experimental evidence implies that
$\thetab < 10^{-9}$, and thus CP is nearly preserved by the strong
interactions. 
The absence of a symmetry to ensure 
$\thetab=0$ is one of the most perplexing features of 
the Standard Model. Three different solutions are commonly considered for 
this ``strong
CP puzzle'': dynamical relaxation of $\thetab$ by means of an axion
field~\cite{Turner:1990vc};
spontaneous CP violation at high energy with a Nelson-Barr
mechanism~\cite{Nelson:1984zb,Barr:1984qx} to ensure reality of the
quark mass determinant; and a vanishing up quark
mass $m_u=0$.  In the particularly simple case of  vanishing $\mup$
the determinant of the quark mass 
matrix is zero, and $\thetab$ is no longer a physical parameter. 
Models in which the up-type quark mass matrix has rank two at short
distance as an accidental consequence of symmetry are easily constructed
\cite{Leurer:1993wg,Leurer:1994gy,Banks:1994yg,Nelson:1997km,Kaplan:1998jk}. 

The viability of a massless up quark can in principle be determined by 
comparing the predictions of QCD with the
observed spectrum and interactions of hadrons.  However, {\em effective}
hadronic theories  whose sole input from 
QCD is the approximate
$SU(3)\times SU(3)$ chiral flavor  symmetry cannot settle this
question.
Under $SU(3)_L\times SU(3)_R$, the
quark mass matrix $\CM$ transforms as the $(3,\bar 3)$ representation, 
and therefore so does the matrix ${\CM^{\dagger}}^{-1}\vert\CM\vert$.  
Thus, symmetry considerations alone can never rule out the
possibility that explicit symmetry breaking occurs in hadronic
physics through combinations of the form 
$\CM_{eff}=\CM +{\CM^{\dagger}}^{-1}\vert\CM \vert/\Lambda$, 
with $\Lambda$ a scale determined by strong interactions, an ambiguity
 pointed out by Kaplan and Manohar \cite{Kaplan:1986ru}.  
In the case $m_u=0$, this gives an effective quark mass matrix
\beq
\CM_{eff} = \(\matrix {m_d^* m_s^*/\Lambda & &\cr & m_d&\cr &&m_s\cr}\)
\eeq
which allows the quantity $m^*_d m^*_s/\Lambda$ to fulfill the role
conventionally played by $m_u$. Note, however, 
that in this case $\det\CM_{eff}$ is
real, and there is no strong CP problem. 
The nonlinear term in $\CM_{eff}$ could
arise from instantons, for example, where the $d$ and $s$ quark
zero-mode  propagators  are connected by $m_d$ and $m_s$ insertions
respectively~\cite{Kaplan:1986ru,Georgi:1981be,Choi:1988sy}. Appropriate
values for the  $\Lambda$ in this scenario would correctly fit all 
hadron data. The conventional extraction of quark mass ratios
from chiral perturbation theory~\cite{Gasser:1982ap} would
nevertheless incorrectly 
yield a nonzero value for $m_u/m_d$.

To resolve the ambiguity chiral symmetry may be supplemented
with additional assumptions.  In Ref.~\cite{Leutwyler:1996et},
Leutwyler gives a list of three plausible assumptions, each of which
independently rules out $m_u=0$.  The assumptions 
are: $SU(3)$ symmetry is always approximately valid, and
all physical quantities can be reliably expanded in powers of $m_s$;
dispersion relations are saturated by the lowest lying states; and
the large-$N_c$ explanation for Zweig's rule (suppression of
virtual quark loops by $1/N_c$) is valid.

We consider the above arguments against $m_u=0$ to be quite
reasonable. Each of these assumptions can be experimentally tested in
various ways, and none has yet been disproven.  Yet it is possible that
an  effective up quark mass, although
sub-leading in  $1/N_c$,  is numerically large enough to account for the
hadron spectrum without implying a general breakdown of chiral
symmetry. Due to the importance of the strong CP problem, we consider it
essential to pin down the value of $m_u$ without  assumptions. 

Lattice simulation of QCD can, in principle, decide whether or not 
a massive up quark is required, particularly in light of recent
advances in realizing chiral symmetry on the 
lattice \cite{Blum:1998ud,Neuberger:1999ry,Luscher:1999mt}. 
With the aid of chiral perturbation theory, 
it is not necessary to  simulate  QCD 
with  a  massless or extremely light quark. 
We begin by reviewing chiral perturbation
theory in the continuum, and then we consider in turn full,
partially quenched, and quenched QCD on the lattice. 
We find that lattice simulations in either full QCD or
partially quenched QCD with all quark masses comparable to the 
strange quark mass could largely 
settle the issue. 

\section{$m_u$ in the continuum}
\label{sec:2}
The light quark masses $(m_u,m_d,m_s)$ are small compared with
the  characteristic 
mass scale of the strong interactions, and QCD possesses an approximate
$SU(3)\otimes SU(3)$ chiral symmetry  broken
spontaneously to the vector $SU(3)$ subgroup, resulting in a light
pseudo-Goldstone boson octet with predictable low energy interactions 
\cite{Gasser:1982ap,Weinberg:1977hb,Weinberg:1979kz,Gasser:1983kx,Gasser:1984yg,Gasser:1985gg}.
For sufficiently light quarks, the pseudo-Goldstone masses squared are
nearly linear in the quark masses, and the quark mass ratios  
may be extracted from the pseudoscalar spectrum with
values $m_u/m_d=0.56$, $m_s/m_d=20.1$ 
\cite{Weinberg:1977hb}\footnote{Similar earlier 
 quark mass estimates were given in Ref.
\cite{Gasser:1975wd}.}.
However, the strange quark mass is large
enough  that the lowest order predictions from chiral symmetry 
receive significant  corrections, of size $m_s/\Lambda_\chi$, 
where $\Lambda_\chi$ is the chiral symmetry breaking
scale of order a GeV. The simplest way of extracting predictions from
chiral symmetry to any given order in $m_s$ and in pion momenta is to use a
phenomenological chiral Lagrangian \cite{Weinberg:1979kz}.  
Using the parametrization of
Gasser and Leutwyler \cite{Gasser:1982ap,Gasser:1983kx,Gasser:1984yg,Gasser:1985gg}
the Lagrangian relevant for the extraction of
quark masses to second order in $m_s$ is
\beq\CL&=&\CL_2+\CL_4+\ldots\\
\CL_2&=&{f^2\over4}\tr\left(\partial_\mu U^\dagger\partial^\mu
  U\right)+{f^2\over2}\tr\left(\chi^\dagger U+\chi U^\dagger\right)\\
\CL_4&=&\ldots+L_4\tr\left(\partial_\mu U^\dagger\partial^\mu
  U\right)\tr\left(U^\dagger\chi+\chi^\dagger U\right)\nn \\
   &&+L_5\tr\left[\partial_\mu U^\dagger\partial^\mu
  U\left(U^\dagger\chi+\chi^\dagger U\right)\right]\nn \\
&&+L_6\left[\tr\left(U^\dagger\chi+\chi^\dagger
    U\right)\right]^2+L_7\left[\tr\left(U^\dagger\chi-\chi^\dagger
    U\right)\right]^2\nn\\ 
&&+L_8\tr\left(\chi^\dagger U\chi^\dagger U+
\chi U^\dagger\chi U^\dagger\right)+\ldots+L_{12}\tr\chi\chi^\dagger
\eeq
where
\beq
\CM&\equiv&\diag(\mup,\mdown,\ms)\\
U&\equiv&\exp\left({i2T_a \pi_a\over f}\right)\\
\chi&\equiv& 2 \CM B\ ,
\eeq
$T_a$ are SU(3) generators, $\pi_a$ are pseudoscalar fields,
and $B,L_i,f$ are phenomenological parameters characterizing QCD
dynamics. (It is not possible to separately determine $\CM$
and $B$ from experiment.) Most of these parameters are renormalization
scale dependent---we follow the usual practice in the chiral
perturbation theory literature of quoting all parameters
at the $\rho$ mass. They are then easily run to a different  scale
by using the renormalization group.
For convenience electromagnetism has been left out of this effective
theory, although electromagnetic effects contribute to the pseudoscalar
meson masses. We therefore define ``QCD'' masses which
have electromagnetic effects subtracted---to leading order these
are just the physical meson masses with the exception of the
electrically charged mesons~\cite{Dashen:1969eg}
\beq
{M_{\pi^\pm}^2}_{QCD}&\approx& {M_{\pi^0}^2}_{\rm phys}\\
{M_{K^\pm}^2}_{QCD}&\approx& {M_{K^\pm}^2}_{\rm
  phys}-{M_{\pi^\pm}^2}_{\rm phys}+{M_{\pi^0}^2}_{\rm phys}\ .
\eeq 
These lowest order  formula for the QCD masses can be improved beyond
leading order
\cite{Donoghue:1993ha,Bijnens:1993ae,Duncan:1996xy,Baur:1996ig,Donoghue:1997zn,Bijnens:1997kk}.
All meson masses used in  subsequent formul\ae\ are these QCD masses.
  From this effective theory, it is possible
to determine a constraint on quark mass ratios, up to
corrections of order $m_s^2$,
\cite{Gasser:1984yg,Gasser:1985gg,Leutwyler:1989pn} 
\beq
{\ms^2-\mh^2\over \mdown^2-\mup^2}\equiv Q^2\approx Q^2_D\equiv{M_K^2\over M_\pi^2}{\left(M^2_{K}-M^2_{\pi}\right)
\over
\left(M^2_{K^0}-M^2_{K^\pm}\right)}\approx(24.2)^2\ ,
\eeq
where
\beq
\mh\equiv&{1\over2}\left(\mup+\mdown\right)\ .
\eeq

If  the values of the  $L$'s are known,  the quark mass ratios may be found:  
\beq
{\ms\over\mh}&=&{2M_K^2\over M_\pi^2(1+\Delta_M)}-1\\
{\mup\over\mh}&=&1-{M_K^2\over M_\pi^4}{M_K^2-(1+\Delta_M)M_\pi^2\over
Q^2(1+\Delta_M)^2}\eeq
where
\beq\Delta_M\equiv-\mu_\pi+\mu_\eta+{8\over
  f^2}(M^2_K-M^2_\pi)(2L_8-L_5)
\eeq
and
\beq
\mu_P\equiv {M_P^2\over32\pi^2f^2}\log\left(M^2_P\over\mu^2\right)\ .
\eeq
To decide whether $m_u$ can be zero,  the value of 
$2L_8-L_5$ is required. $L_5$ can be determined
from the ratio $f_K/f_\pi$,  with a value
$10^3L_5\approx1.4\pm0.5$. The linear combination $L_5-12 L_7-6 L_8$
can be 
extracted from the  pseudoscalar masses, giving
$10^3(2L_7+L_8)=\pm0.2$.
(The quoted errors represent the estimated theoretical
uncertainty due to
higher order corrections in $m_s$.) In principle (but not in practice)
$L_4$ and  $L_6-L_7$ can be determined from meson-meson interactions.
However  data cannot be used {\em even in principle} to completely specify
the $L$'s.
The Gasser-Leutwyler Lagrangian is invariant
under the replacement~\cite{Gasser:1985gg}
\beq \chi&\rightarrow& \chi+\beta(\det \chi){\chi}^{-1}\\
L_6\rightarrow L_6-\delta\quad L_7&\rightarrow&L_7-\delta\quad
L_8\rightarrow L_8+2\delta\eeq
with \beq\delta\equiv\beta f^2/32\ .\eeq This invariance is a
necessary consequence of the Kaplan-Manohar ambiguity, and leads to a
corresponding ambiguity in the $L$'s.

A combination of phenomenological and large $N_c$
constraints {\em can} be used to determine all the $L$'s
with an estimated 30\%
precision~\cite{Leutwyler:1989pn,Ecker:1989yg,Ecker:1989te}.  
The results  agree (by design) with the picture that all
terms in $\CL_4$ are consistent with the values obtained by
integrating out light resonances, and are  inconsistent with $m_u=0$
\cite{Leutwyler:1989pn}. 
In particular,
a conventional estimate of this kind
gives~\cite{Pich:1993uq,Bijnens:1994qh,Donoghue:1994nz,Ecker:1995gg}
$10^3L_8\approx0.9\pm0.3$, $10^3L_7\approx (-0.4\pm0.2)$,
while the hypothesis $m_u=0$ would require
$10^3L_8=-0.4\pm0.3$, and $10^3L_7= 0.2\pm0.2$. 
Thus any method which
can directly determine the $L$'s, even with  errors  as large as 100\%,
can distinguish between these two possibilities.

\section{$m_u$ on the lattice}
\label{sec:3}

Simulating QCD on a lattice is, in principle, the most reliable way of
determining quantities which are not predicted from 
symmetry alone. For instance, although it is only possible to
constrain ratios of light
quark masses phenomenologically, recent lattice computations of the
hadron spectrum  have suggested an absolute range for the strange quark
mass
\cite{Bhattacharya:1997ht,Gupta:1997sa,Gough:1997kw,Mackenzie:1997sd,Gimenez:1998uv,Gupta:1998bm,Lubicz:1998kc,Capitani:1998mq,Garden:1999fg,Aoki:1999mr}
and the parameter $B$.  It is difficult to simulate  quarks 
with realistic masses: light quarks require large  lattices 
to avoid finite volume artifacts. However with the aid of
chiral 
symmetry lattice computations done at
moderate quark masses can be extrapolated to lighter quark masses.
Unfortunately this 
extrapolation is complicated  and plagued with unphysical 
artifacts in the usual quenched approximation, where
quark dynamics are not included.

\subsection{Unquenched lattice QCD}
\label{sec:3a}
With computing power sufficient to simulate unquenched QCD with 
quark masses light enough to apply chiral symmetry, 
there are several possible methods\footnote{For earlier work on
lattice computations of chiral coefficients 
see refs. \cite{Myint:1994yw,Levi:1997ur}.}
to extract  the chiral Lagrangian coefficients and determine $m_u$.
One way is to use the chiral symmetry prediction for the pion
mass squared \cite{Gasser:1985gg}:

\beq
M_\pi^2=2\mh B\left(1+\mu_\pi-{1\over3}\mu_\eta+2\mh
  K_3+{m_u+m_d+m_s\over 3}K_4\right)
\eeq
where
\beq
K_3&\equiv&{8 B\over f^2}(2L_8-L_5)\\
K_4&\equiv&{48 B\over f^2}(2 L_6-L_4) \ .
\eeq

By measuring the pion mass as a function of $\mh$ and $m_s$
the combination $2L_8-L_5$, needed in determining the quark mass
ratios,  
as well as the combination $2 L_6-L_4$, which  provides an interesting test
of conventional assumptions and the large $N_c$ expansion, may be
extracted. 
Independently varying $\mh$ and $m_s$ requires a lot of different
simulations, however. It may be simpler to work with equal quark
masses $m_u=m_d=m_s=\mh=\mt$ and vary $\mt$. A fit to the quadratic
dependence of the pion mass squared then yields the linear combination 
$2L_8-L_5+6L_6-3L_4$ from
\beq{\partial^2M_\pi^2\over \partial \mt^2}={B^2\over
    12\pi^2f^2}\left[3+2\log\left({2 \mt
        B\over\mu^2}\right)+768\pi^2(2L_8-L_5+6L_6-3L_4)\right]\ .
\eeq 
With equal quark masses, the combination $2L_6-L_4$ may be separately
extracted by a 
measurement of the matrix element of $\bar s s$ in the pion, using
\beq\langle\pi|\bar s s |\pi\rangle=-{\mt B^2\over 36 \pi^2 f^2}\left[
   1+\log\left(2\mt B\over\mu^2\right)+1152\pi^2(L_4-2L_6)\right]\ .
\eeq
$L_6$ may
be independently extracted from the vacuum expectation value 
$\langle\int\bar s s \int\bar d d\rangle$, and $L_4$ may also be measured from the
dependence of $f_\pi$ on $\mt$. 
Thus in principle  by simulating QCD on the lattice with several
different quark masses
it is possible to verify the conventional estimates of
those $L$'s for which there is no direct experimental data. Even a
measurement of $2L_8-L_5$ with  100\% errors provides an interesting 
test of the large $N_c$ expansion and, depending on the result, 
could rule out the possibility that $m_u=0$. Note that  measurement 
of $L_4$ and
$L_6$ would also be quite  interesting theoretically, although not directly 
needed to extract quark mass ratios.  Available data provides no
constraints on $L_{4,6}$. 

Ruling out $\mup=0$ along the above lines may not be easy.
Here we define the more restrictive ``effective up mass
hypothesis,'' which may be somewhat simpler to test than whether
$m_u=0$, since this hypothesis makes a prediction for $L_6$. The
effective up mass
hypothesis is motivated by the agreement between different
experimental determinations of light quark mass 
ratios \cite{Gasser:1982ap,Leutwyler:1989pn,Leutwyler:1994pf,Leutwyler:1996qg,Leutwyler:1996sa}.
The hypothesis is that $\mup=0$, but that the conventional
low energy theory works accurately with the replacement
$\mup\to m^*_d m^*_s/\Lambda$. In the chiral lagrangian,  such an
effective up mass is a nonstandard contribution to the coefficients 
$L_6, L_7,$ and
$L_8$, in the combination $\Delta L_6$=$\Delta L_7=-2\Delta L_8=0.7\times10^{-3}$. 
With this hypothesis, all the $L$'s are determined, and
$L_4$  agrees with the conventional estimate $10^3L_4=-0.3\pm
0.5$, while $10^3L_6$ is $0.5\pm0.3$, in contrast to the 
conventional estimate $10^3L_6=-0.2\pm0.3$. 

\subsection{Partially quenched lattice QCD}
\label{sec:3b}
QCD simulations in
the ``partially quenched''approximation, which includes the effects of
N flavors of dynamical quarks with mass $\msea$ different
from the valence quark mass, may also be of interest.  Such
simulations could 
be decisive for the determination of the Gasser-Leutwyler $L$ 
coefficients\footnote{After the completion
of this work we were informed of the work of Sharpe and
Shoresh~\cite{Sharpe} who reach similar conclusions.}. 
 For such an analysis to be reliable it is necessary that the sea
quark mass is sufficiently small  for the dominant artifacts of partial
quenching  to be computable using partially quenched chiral perturbation
theory~\cite{Sharpe:1997by}. This requires that the mass $M_{SS}$ of a
pion made of 
sea quarks be light compared with the scale $\Lambda_\chi$. It is also
desirable to take the valence quark mass to be comparable to the sea
quark mass, to reduce quenching artifacts from the non-decoupling of
the $\eta'$ \cite{Golterman:1998st}\footnote{Note added in revision: Recent work by Sharpe and Shoresh \cite{SS} has shown that the artifacts from the $\eta'$ are under theoretical control even when the valence quark mass is much lighter than the sea quark mass, provided both masses are sufficiently small.}. In the limit
$\msea\sim\mt\ll\Lambda_\chi$, 
Sharpe has calculated the following dependence of the pion  masses on
the valence 
and sea quark masses  \cite{Sharpe:1997by}:
\beq M_{\pi^\pm}^2&=&2\mt B\Big\{ 1+
{B\over N 8\pi^2f^2}\Big[(2\mt-\msea)\log\big({2\mt B\over\mu^2}\big)
+\mt -\msea  \Big]\nn \\
&&+{16\mt B\over f^2}(2L_8-L_5)+{N 16\msea B\over f^2}(2L_6-L_4)\Big\}\\
M_{SV}^2&=&(\msea+\mt) B\Big[ 1+
{\mt B\over N 8\pi^2 f^2}\log\left({2\mt B\over\mu^2}\right) 
+{8(\msea+\mt) B\over f^2}(2L_8-L_5)\nn\\
&&+{N 16\msea B\over
f^2}(2L_6-L_4)\Big]\\
M_{SS}^2&=&2\msea B\Big[ 1+
{\msea B\over N 8\pi^2f^2}\log\left({2\msea B\over\mu^2}\right)\nn \\ 
&&+{16\msea B\over f^2}(2L_8-L_5)+{N 16\msea B\over
f^2}(2L_6-L_4)\Big]\ .\eeq
Here all valence quarks have mass $\mt$, $\pi^\pm$ is a pion made of
different valence quarks, $M_{SS}$ is the mass of a pion made of sea quarks and
$M_{SV}$ is the mass of a pion made of one sea and one valence
quark. For $N=3$ the parameters $B$ and $L_i$ are the same as
those in the QCD chiral lagrangian
\cite{Sharpe:1997by,Bernard:1994sv}. 
Thus for $N=3$,
 the desired combination $2L_8-L_5$
may be extracted 
 by fitting the pion masses as
a function of $\mt$ with $\msea$ held fixed.   Note that  an $N=2$
simulation, while interesting, is not
sufficient to determine the $L$ coefficients, as these  may have significant 
dependence on
the number of flavors. For instance the contributions  
from gauge field configurations with fermion zero modes, such as
instantons,  should be quite sensitive to the number of sea flavors.
 
Golterman and Leung \cite{Golterman:1998st} have  extended
the partially quenched   
chiral perturbation theory calculations to the case
where the $\eta'$ is light compared to  the 
scale $\Lambda_\chi$, as expected in the large $N_c$ limit.
In this limit, unless the valence and sea quark pion  masses are
comparable and both much lighter than the
$\eta'$,  the pion
masses depend on two new parameters associated with the $\eta'$ mass
and decay constant. 
Even with a light $\eta'$ it is  theoretically possible,  with enough
different measurements, to extract $2L_8-L_5$ from lattice data\footnote{Note added in revision: see recent work of Sharpe and Shoresh \cite{SS} for how to deal with the $\eta'$ artifacts.}.

\subsection{Quenched lattice QCD}
\label{sec:3c}
The quenched approximation is not a systematic approximation to QCD. 
Nevertheless it
is generally used to facilitate lattice computations with currently
available computing power, and quenched QCD
does have a
spectrum similar to the real thing \cite{Kanaya:1998sd,Aoki:1999yr}. 
This suggests that the
dominant effects of quark loops can be compensated for in the
quenched approximation by
adjusting  the  QCD parameters (quark masses and the scale
$\Lambda_{QCD}$).  
However a reliable extraction of the true value of the chiral Lagrangian
coefficients and  of $\mup$ from the
quenched spectrum is problematic; 
the possibility that  the quenched
approximation with a nonzero up quark mass mimics the true QCD
spectrum with $\mup=0$ cannot be ruled out. For instance the 
quenched approximation is missing the down and strange 
quark loop effects which, in
conjunction with instantons, might mimic an effective up quark mass
 proportional to $m^*_d m^*_s$.
 
A possible way to explore the quark loop contribution to the
effective up quark mass in the quenched approximation is to explicitly include
sources for the sea quarks. For instance,  one could measure a
three-point function 
\beq
\label{eq:dominatrix}
\langle\pi|\int
\bar s s|\pi\rangle \ .
\eeq 
In particular, the instanton effects which might give an effective up
quark mass do  contribute to this matrix element in the quenched approximation.
 In full QCD this matrix element  is equivalent
to $\partial M^2_\pi/\partial m_s$, which would vanish ignoring quark
loops; however the equivalence does not hold in the quenched
approximation. Logarithmically enhanced quenched 
artifacts from an $\eta'$ loop\footnote{We thank
Steve Sharpe for explaining this to us.} give a contribution to this matrix
element which are suppressed by $1/N_c^2$ but which introduce a new parameter
which cannot be computed using quenched chiral perturbation theory
\cite{Sharpe:1992ft,Bernard:1992mk}. To avoid this 
artifact
it may be better to measure 
\beq
\label{eq:dominatrixII}
\langle\pi|\int
(2\bar s s-\bar d d-\bar u u)|\pi\rangle \ .
\eeq  
 
\section{Summary}
Chiral perturbation theory makes possible the use of
lattice simulations of full and partially quenched QCD with moderate
quark masses to learn about the properties of QCD with light
quarks. In this paper we showed how to use full or partially quenched
simulations  with equal, moderately sized  quark masses to extract the
second order coefficients in the pion chiral Lagrangian.  Such
calculations are of interest to check the predictions of large $N_c$
QCD, to verify
chiral perturbation theory, and to test the hypothesis of resonance
saturation of dispersion relations. Such computations, even with
large errors, can provide a
method for settling the important issue of whether the $\mup=0$
solution to the strong CP problem is consistent with the spectrum of light
pseudoscalar mesons.

\bigskip
\noindent\medskip\centerline{\bf Acknowledgments}

We thank Steve Sharpe for informing us, 
prior to publication, of his independent  work with
Noam Shoresh  on partially quenched extraction of the chiral Lagrangian
coefficients~\cite{Sharpe}, and for useful comments.
We also thank the hospitality of the CERN theory group, where this work
was initiated, and of the Aspen Center for Physics, where this work
was completed. A.G.C. is supported in part by DOE grant
\#DE-FG02-91ER-40676; D.B.K. is supported in part by DOE grant
\#DOE-ER-40561;
A.E.N. is supported in part by DOE grant \#DE-FG03-96ER40956.

\bibliography{quark}

\providecommand{\href}[2]{#2}\begingroup\raggedright\begin{thebibliography}{10}

\bibitem{Turner:1990vc}
M.~S. Turner, ``Windows On The Axion,'' {\em Phys. Rept.} {\bf 197} (1990)
  67--97.

\bibitem{Nelson:1984zb}
A.~Nelson, ``Naturally Weak CP Violation,'' {\em Phys. Lett.} {\bf 136B} (1984)
  387.

\bibitem{Barr:1984qx}
S.~M. Barr, ``Solving the Strong CP Problem Without the Peccei-Quinn
  Symmetry,'' {\em Phys. Rev. Lett.} {\bf 53} (1984) 329.

\bibitem{Leurer:1993wg}
M.~Leurer, Y.~Nir, and N.~Seiberg, ``Mass matrix models,'' {\em Nucl. Phys.}
  {\bf B398} (1993) 319--342,
  \href{http://xxx.lanl.gov/abs/hep-ph/9212278}{{\tt hep-ph/9212278}}.

\bibitem{Leurer:1994gy}
M.~Leurer, Y.~Nir, and N.~Seiberg, ``Mass matrix models: The Sequel,'' {\em
  Nucl. Phys.} {\bf B420} (1994) 468--504,
  \href{http://xxx.lanl.gov/abs/hep-ph/9310320}{{\tt hep-ph/9310320}}.

\bibitem{Banks:1994yg}
T.~Banks, Y.~Nir, and N.~Seiberg, ``Missing (up) mass, accidental anomalous
  symmetries, and the strong CP problem,''
  \href{http://xxx.lanl.gov/abs/hep-ph/9403203}{{\tt hep-ph/9403203}}.

\bibitem{Nelson:1997km}
A.~E. Nelson and M.~J. Strassler, ``A Realistic supersymmetric model with
  composite quarks,'' {\em Phys. Rev.} {\bf D56} (1997) 4226--4237,
  \href{http://xxx.lanl.gov/abs/hep-ph/9607362}{{\tt hep-ph/9607362}}.

\bibitem{Kaplan:1998jk}
D.~E. Kaplan, F.~Lepeintre, A.~Masiero, A.~E. Nelson, and A.~Riotto, ``Fermion
  masses and gauge mediated supersymmetry breaking from a single U(1),''
  \href{http://xxx.lanl.gov/abs/hep-ph/9806430}{{\tt hep-ph/9806430}}.

\bibitem{Kaplan:1986ru}
D.~B. Kaplan and A.~V. Manohar, ``Current Mass Ratios of the Light Quarks,''
  {\em Phys. Rev. Lett.} {\bf 56} (1986) 2004.

\bibitem{Georgi:1981be}
H.~Georgi and I.~N. McArthur, ``Instantons And The Up Quark Mass,''.
  HUTP-81/A011.

\bibitem{Choi:1988sy}
K.~Choi, C.~W. Kim, and W.~K. Sze, ``Mass Renormalization By Instantons And The
  Strong CP Problem,'' {\em Phys. Rev. Lett.} {\bf 61} (1988) 794.

\bibitem{Gasser:1982ap}
J.~Gasser and H.~Leutwyler, ``Quark Masses,'' {\em Phys. Rept.} {\bf 87} (1982)
  77--169.

\bibitem{Leutwyler:1996et}
H.~Leutwyler, ``Light quark effective theory,''
  \href{http://xxx.lanl.gov/abs/hep-ph/9609465}{{\tt hep-ph/9609465}}.

\bibitem{Blum:1998ud}
T.~Blum, ``Domain wall fermions in vector gauge theories,'' {\em Nucl. Phys.
  Proc. Suppl.} {\bf 73} (1999) 167,
  \href{http://xxx.lanl.gov/abs/hep-lat/9810017}{{\tt hep-lat/9810017}}.

\bibitem{Neuberger:1999ry}
H.~Neuberger, ``Chiral fermions on the lattice,''
  \href{http://xxx.lanl.gov/abs/hep-lat/9909042}{{\tt hep-lat/9909042}}.

\bibitem{Luscher:1999mt}
M.~Luscher, ``Chiral gauge theories on the lattice with exact gauge
  invariance,'' \href{http://xxx.lanl.gov/abs/hep-lat/9909150}{{\tt
  hep-lat/9909150}}.

\bibitem{Weinberg:1977hb}
S.~Weinberg, ``The Problem Of Mass,'' {\em Trans. New York Acad. Sci.} {\bf 38}
  (1977) 185--201.

\bibitem{Weinberg:1979kz}
S.~Weinberg, ``Phenomenological Lagrangians,'' {\em Physica} {\bf 96A} (1979)
  327.

\bibitem{Gasser:1983kx}
J.~Gasser and H.~Leutwyler, ``Low-Energy Theorems As Precision Tests Of Qcd,''
  {\em Phys. Lett.} {\bf 125B} (1983) 325.

\bibitem{Gasser:1984yg}
J.~Gasser and H.~Leutwyler, ``Chiral Perturbation Theory To One Loop,'' {\em
  Ann. Phys.} {\bf 158} (1984) 142.

\bibitem{Gasser:1985gg}
J.~Gasser and H.~Leutwyler, ``Chiral Perturbation Theory: Expansions In The
  Mass Of The Strange Quark,'' {\em Nucl. Phys.} {\bf B250} (1985) 465.

\bibitem{Gasser:1975wd}
J.~Gasser and H.~Leutwyler, ``Implications Of Scaling For The Proton - Neutron
  Mass - Difference,'' {\em Nucl. Phys.} {\bf B94} (1975) 269.

\bibitem{Dashen:1969eg}
R.~Dashen, ``Chiral SU(3) x SU(3) as a symmetry of the strong interactions,''
  {\em Phys. Rev.} {\bf 183} (1969) 1245--1260.

\bibitem{Donoghue:1993ha}
J.~F. Donoghue, B.~R. Holstein, and D.~Wyler, ``Mass ratios of the light
  quarks,'' {\em Phys. Rev. Lett.} {\bf 69} (1992) 3444.

\bibitem{Bijnens:1993ae}
J.~Bijnens, ``Violations of Dashen's theorem,'' {\em Phys. Lett.} {\bf B306}
  (1993) 343--349, \href{http://xxx.lanl.gov/abs/hep-ph/9302217}{{\tt
  hep-ph/9302217}}.

\bibitem{Duncan:1996xy}
A.~Duncan, E.~Eichten, and H.~Thacker, ``Electromagnetic splittings and light
  quark masses in lattice QCD,'' {\em Phys. Rev. Lett.} {\bf 76} (1996)
  3894--3897, \href{http://xxx.lanl.gov/abs/hep-lat/9602005}{{\tt
  hep-lat/9602005}}.

\bibitem{Baur:1996ig}
R.~Baur and R.~Urech, ``On the corrections to Dashen's theorem,'' {\em Phys.
  Rev.} {\bf D53} (1996) 6552--6557,
  \href{http://xxx.lanl.gov/abs/hep-ph/9508393}{{\tt hep-ph/9508393}}.

\bibitem{Donoghue:1997zn}
J.~F. Donoghue and A.~F. Perez, ``The Electromagnetic mass differences of pions
  and kaons,'' {\em Phys. Rev.} {\bf D55} (1997) 7075--7092,
  \href{http://xxx.lanl.gov/abs/hep-ph/9611331}{{\tt hep-ph/9611331}}.

\bibitem{Bijnens:1997kk}
J.~Bijnens and J.~Prades, ``Electromagnetic corrections for pions and kaons:
  Masses and polarizabilities,'' {\em Nucl. Phys.} {\bf B490} (1997) 239--271,
  \href{http://xxx.lanl.gov/abs/hep-ph/9610360}{{\tt hep-ph/9610360}}.

\bibitem{Leutwyler:1989pn}
H.~Leutwyler, ``How About m(u) = 0?,'' {\em Nucl. Phys.} {\bf B337} (1990) 108.

\bibitem{Ecker:1989yg}
G.~Ecker, J.~Gasser, H.~Leutwyler, A.~Pich, and E.~de~Rafael, ``Chiral
  Lagrangians For Massive Spin 1 Fields,'' {\em Phys. Lett.} {\bf B223} (1989)
  425.

\bibitem{Ecker:1989te}
G.~Ecker, J.~Gasser, A.~Pich, and E.~de~Rafael, ``The Role Of Resonances In
  Chiral Perturbation Theory,'' {\em Nucl. Phys.} {\bf B321} (1989) 311.

\bibitem{Pich:1993uq}
A.~Pich, ``Introduction To Chiral Perturbation Theory,''
  \href{http://xxx.lanl.gov/abs/hep-ph/9308351}{{\tt hep-ph/9308351}}.

\bibitem{Bijnens:1994qh}
J.~Bijnens, G.~Ecker, and J.~Gasser, ``Chiral perturbation theory,''
  \href{http://xxx.lanl.gov/abs/hep-ph/9411232}{{\tt hep-ph/9411232}}.

\bibitem{Donoghue:1994nz}
J.~F. Donoghue, ``Light quark masses and mixing angles,''
  \href{http://xxx.lanl.gov/abs/hep-ph/9403263}{{\tt hep-ph/9403263}}.

\bibitem{Ecker:1995gg}
G.~Ecker, ``Chiral Perturbation Theory,'' {\em Prog. Part. Nucl. Phys.} {\bf
  35} (1995) 1--80, \href{http://xxx.lanl.gov/abs/hep-ph/9501357}{{\tt
  hep-ph/9501357}}.

\bibitem{Bhattacharya:1997ht}
T.~Bhattacharya and R.~Gupta, ``Advances in the determination of quark
  masses,'' {\em Nucl. Phys. Proc. Suppl.} {\bf 63} (1998) 95,
  \href{http://xxx.lanl.gov/abs/hep-lat/9710095}{{\tt hep-lat/9710095}}.

\bibitem{Gupta:1997sa}
R.~Gupta and T.~Bhattacharya, ``Light quark masses from lattice QCD,'' {\em
  Phys. Rev.} {\bf D55} (1997) 7203--7217,
  \href{http://xxx.lanl.gov/abs/hep-lat/9605039}{{\tt hep-lat/9605039}}.

\bibitem{Gough:1997kw}
B.~J. Gough {\em et.~al.}, ``The Light quark masses from lattice gauge
  theory,'' {\em Phys. Rev. Lett.} {\bf 79} (1997) 1622--1625,
  \href{http://xxx.lanl.gov/abs/hep-ph/9610223}{{\tt hep-ph/9610223}}.

\bibitem{Mackenzie:1997sd}
P.~B. Mackenzie, ``Recent lattice results on the light quark masses,'' {\em
  Nucl. Phys. Proc. Suppl.} {\bf 53} (1997) 23--29,
  \href{http://xxx.lanl.gov/abs/hep-ph/9609261}{{\tt hep-ph/9609261}}.

\bibitem{Gimenez:1998uv}
V.~Gimenez, L.~Giusti, F.~Rapuano, and M.~Talevi, ``Lattice quark masses: A
  Nonperturbative measure,'' {\em Nucl. Phys.} {\bf B540} (1999) 472,
  \href{http://xxx.lanl.gov/abs/hep-lat/9801028}{{\tt hep-lat/9801028}}.

\bibitem{Gupta:1998bm}
R.~Gupta, ``Quark masses, B parameters, and CP violation parameters epsilon and
  epsilon-prime / epsilon,'' \href{http://xxx.lanl.gov/abs/hep-ph/9801412}{{\tt
  hep-ph/9801412}}.

\bibitem{Lubicz:1998kc}
V.~Lubicz, ``Light quark masses and CKM matrix elements from lattice QCD,''
  {\em Nucl. Phys. Proc. Suppl.} {\bf 74} (1999) 291,
  \href{http://xxx.lanl.gov/abs/hep-ph/9809417}{{\tt hep-ph/9809417}}.

\bibitem{Capitani:1998mq}
S.~Capitani, M.~Luscher, R.~Sommer, and H.~Wittig, ``Non-perturbative quark
  mass renormalization in quenched lattice QCD,'' {\em Nucl. Phys.} {\bf B544}
  (1999) 669, \href{http://xxx.lanl.gov/abs/hep-lat/9810063}{{\tt
  hep-lat/9810063}}.

\bibitem{Garden:1999fg}
J.~Garden, J.~Heitger, R.~Sommer, and H.~Wittig, ``Precision computation of the
  strange quark's mass in quenched QCD,''
  \href{http://xxx.lanl.gov/abs/hep-lat/9906013}{{\tt hep-lat/9906013}}.

\bibitem{Aoki:1999mr}
{\bf JLQCD} Collaboration, S.~Aoki {\em et.~al.}, ``Nonperturbative
  determination of quark masses in quenched lattice QCD with the Kogut-Susskind
  fermion action,'' {\em Phys. Rev. Lett.} {\bf 82} (1999) 4392--4395,
  \href{http://xxx.lanl.gov/abs/hep-lat/9901019}{{\tt hep-lat/9901019}}.

\bibitem{Myint:1994yw}
S.~Myint and C.~Rebbi, ``Derivation of chiral Lagrangians from lattice QCD,''.
  Prepared for LATTICE 93: 11th International Symposium on Lattice Field
  Theory, Dallas, TX, 12-16 Oct 1993.

\bibitem{Levi:1997ur}
A.~R. Levi, V.~Lubicz, and C.~Rebbi, ``Towards a lattice calculation of the
  coefficients of the QCD chiral Lagrangian,'' {\em Nucl. Phys. Proc. Suppl.}
  {\bf 53} (1997) 275--277, \href{http://xxx.lanl.gov/abs/hep-lat/9607025}{{\tt
  hep-lat/9607025}}.

\bibitem{Leutwyler:1994pf}
H.~Leutwyler, ``Masses of the light quarks,''
  \href{http://xxx.lanl.gov/abs/hep-ph/9405330}{{\tt hep-ph/9405330}}.

\bibitem{Leutwyler:1996qg}
H.~Leutwyler, ``The Ratios of the Light Quark Masses,'' {\em Phys. Lett.} {\bf
  B378} (1996) 313--318, \href{http://xxx.lanl.gov/abs/hep-ph/9602366}{{\tt
  hep-ph/9602366}}.

\bibitem{Leutwyler:1996sa}
H.~Leutwyler, ``Bounds on the light quark masses,'' {\em Phys. Lett.} {\bf
  B374} (1996) 163--168, \href{http://xxx.lanl.gov/abs/hep-ph/9601234}{{\tt
  hep-ph/9601234}}.


\bibitem{Sharpe}
S.~Sharpe and N.~Shoresh,
``Partially quenched QCD with non-degenerate dynamical quarks,''
Nucl.\ Phys.\ Proc.\ Suppl.\  {\bf 83} (2000)  968. 
{\tt hep-lat/9909090}.


\bibitem{SS}
S.~Sharpe and N.~Shoresh,
``Physical results from unphysical simulations,''
Phys.\ Rev. {\bf D62} (2000) 094503. 
{\tt hep-lat/0006017}.

\bibitem{Sharpe:1997by}
S.~R. Sharpe, ``Enhanced chiral logarithms in partially quenched QCD,'' {\em
  Phys. Rev.} {\bf D56} (1997) 7052--7058,
  \href{http://xxx.lanl.gov/abs/hep-lat/9707018}{{\tt hep-lat/9707018}}.

\bibitem{Golterman:1998st}
M.~F.~L. Golterman and K.-C. Leung, ``Applications of partially quenched chiral
  perturbation theory,'' {\em Phys. Rev.} {\bf D57} (1998) 5703--5710,
  \href{http://xxx.lanl.gov/abs/hep-lat/9711033}{{\tt hep-lat/9711033}}.

\bibitem{Bernard:1994sv}
C.~W. Bernard and M.~F.~L. Golterman, ``Partially quenched gauge theories and
  an application to staggered fermions,'' {\em Phys. Rev.} {\bf D49} (1994)
  486--494, \href{http://xxx.lanl.gov/abs/hep-lat/9306005}{{\tt
  hep-lat/9306005}}.

\bibitem{Kanaya:1998sd}
{\bf CP-PACS} Collaboration, K.~Kanaya {\em et.~al.}, ``Quenched light hadron
  spectrum with the Wilson quark action: Final results from CP-PACS,'' {\em
  Nucl. Phys. Proc. Suppl.} {\bf 73} (1999) 189,
  \href{http://xxx.lanl.gov/abs/hep-lat/9809146}{{\tt hep-lat/9809146}}.

\bibitem{Aoki:1999yr}
{\bf CP-PACS} Collaboration, S.~Aoki {\em et.~al.}, ``Quenched light hadron
  spectrum,'' \href{http://xxx.lanl.gov/abs/hep-lat/9904012}{{\tt
  hep-lat/9904012}}.

\bibitem{Sharpe:1992ft}
S.~R. Sharpe, ``Quenched chiral logarithms,'' {\em Phys. Rev.} {\bf D46} (1992)
  3146--3168, \href{http://xxx.lanl.gov/abs/hep-lat/9205020}{{\tt
  hep-lat/9205020}}.

\bibitem{Bernard:1992mk}
C.~W. Bernard and M.~F.~L. Golterman, ``Chiral perturbation theory for the
  quenched approximation of QCD,'' {\em Phys. Rev.} {\bf D46} (1992) 853--857,
  \href{http://xxx.lanl.gov/abs/hep-lat/9204007}{{\tt hep-lat/9204007}}.

\end{thebibliography}\endgroup
\bibliographystyle{utcaps}

\end{document}